\documentstyle[12pt]{article}
\begin{document}
\begin{center}
\large {\bf NON-ABELIAN GAUGED CHIRAL BOSON WITH A GENERALIZED FADDEEVIAN REGULARIZATION}\\
\vspace{1.5cm}
Anisur Rahaman\footnote {e-mail: anisur.rahman@saha.ac.in}\\
Durgapur Govt. College, Durgapur-713214, Burdwan,\\
 West Bengal,
INDIA\\
\end{center}

\begin{center}Abstract\end{center}
We consider non-Abelian  gauged version of chiral boson with a generalized 
Faddeevian regularization. It is a second class constrained theory. We quantize 
the theory and analyze the phase space. It is shown that  in spite of the lack 
of 
manifest Lorentz 
invariance in the action, it has a consistent and Poincare' invariant 
phase space structure.
\newpage
\setcounter{page}{1}

Chiral boson has not only acquired a significant position in in the context of
string theory but also it is of particular interest in the context of two
dimensional anomalous gauge theory [1-10]. Two equivalent description are 
available 
for the chiral boson. A manifestly Lorentz invariant formulation was done
by Siegel using an auxiliary field \cite{WS}.  In this formulation it was 
found that 
the classical reparametrization invariance of the action did not maintain
because of the gravitational anomaly at the quantum mechanical level. An 
alternative formulation was given by Floreanini and Jackiw \cite{FJ}. 
Though it did not suffer from Gravitational anomaly it was lacking Lorentz
invariance at the action level however Lorentz invariance was not really broken
there. Subsequently a Lorentz invariant formulation was revived by 
Bellucci,
Golterman  and  Petcher \cite{BEL}.

The
interaction of chiral boson with gauge field has also been extensively 
studied.
The vector way of gauging was found to be inconsistent by Bellucci, Golterman and
Petcher and by
Floreanini and Jackiw \cite{FJ, BEL}. However Chiral way of gauging was found
to render consistent anomalous field theory \cite{RJ, MG, FAD}. 

In Ref \cite{MG}, Mitra and Ghosh presented a new way of gauging Chiral boson. 
The lagrangian 
of this model can be obtained from the Chiral Schwinger model when it is studied
with a Faddeevian regularization \cite{FAD} imposing a chiral constraint 
\cite{KH}. 
The regularization presented by Mitra was improved by Abreu {\it etal}.
\cite{WOT}
and showed 
that in place of a
specific mass like counter term a one parameter class of mass like counter term of 
Faddeevian class lead to a consistent anomalous gauge field theory. Recently
we have presented a consistent description of the above model in terms of chiral 
boson \cite{AR} in a similar way as 
Harada did it in his paper \cite{KH} for the Chiral Schwinger model with  
Jackiw-Rajaraman 
regularization.

In the above models the interacting fields were Abelian type. Various authors
tried to generalize the model introducing the non-Abelian gauge field 
\cite{RAJ, FAD1, ABD, KH1, AS, TUS}. Non-
Abelian generalization of the chiral Schwinger model \cite{RAJ} was appeared immediately
after Jackiw and Rajaraman presented the Abelian version of it 
\cite{RJ}. And later generalization of both the anomalous and non-anomalous
model and also other variants came to the literature \cite{RAJ, FAD1, ABD, KH1, AS, TUS}.  
Faddeev in his paper presented a special type of regularization where Gauss law of the theory
itself becomes second class. However it would be possible to quantize the 
theory. Commonly it is known as Faddeevian regularization. 
Faddeevian class of anomalous gauge theory (regularization) has been becoming 
 a subject of 
considerable interest
for the last few years \cite{MG, WOT, AR}. So the non-Abelian generalization of Chiral Schwinger
model with a generalized Faddeevian Regularization and description of this model
in terms of chiral boson would certainly be a subject of new investigation.

In this letter we have started with a bosonized non-Abelian version of the 
chiral
Schwinger model where Faddeevian class of anomaly appeared because of 
the presence
of generalized regularization proposed by Abreu {\it etal.} in \cite{WOT} 
and reproduce a lagrangian 
which has one more (chiral) constraint in a similar way as it has been done
in \cite{AR} for the Abelian case. To be more specific and in a simple language
we reproduce a 
non-Abelian version of the model  studied in \cite{AR}.
We analyze the phases pace and show that it leads to a consistent and physically sensible anomalous  gauge
theory. We also check the Poincare algebra to ensure the Poincare symmetry because
the lagrangian is lacking Lorentz covariance to start with.

The fermionic form of the non-Abelian theory is described by the generating 
functional  
\begin{equation}
Z=\int{\cal D}A_\mu{\cal D}\bar\psi{\cal D}\psi 
e^{i\int d^2x\bar\psi(i\partial\!\!\!/ - eA\!\!\!/ {(1+i\gamma_5)\over 2})\psi 
- {1\over 4}tr(F_{\mu\nu} F^{\mu\nu}) },
\end{equation}
where $\psi$ is the $(1+1)$ dimensional fermionic field furnishing some representation 
$\Gamma$ of some simple group $G$. It interacts with the gauge field through a chiral 
coupling. $A_\mu$ can be written as $A_\mu^a\tau^a$ with $\tau^a$ being the 
the group generators in $\Gamma$ satisfying $[\tau^a, \tau^b] =
-if^{abc}\tau^c$ and $tr(\tau^a\tau^b) = \delta^{ab}$.
The non-Abelian field strength is
\begin{equation}
F_{\mu\nu}= \partial_\mu A_\nu - \partial_\nu A_\mu - e[A_\mu, A_\nu].
\end{equation}
We use the convention for the  metric $g_{\mu\nu}=diag(+1,-1)$, and for the 
antisymmetric tensor 
$\epsilon^{01}=1$.

Since in the bosonized formulation the anomaly shows up at the classical 
level it is the natural setting for the phase space analysis of this theory.
In order to get the bosonized action we first integrate the fermions out and
construct the effective action. That will yield a determinant of the chiral 
Dirac operator in presence of the background field. So the partition function
becomes
\begin{equation}
Z=\int {\cal D}A_\mu ~~|\det{i({D\!\!\!/}_{ch}[A])}| ~~e^{i\int -{1\over 4}tr(F^2)}.
\end{equation}
The chiral Dirac operator is given by
\begin{equation} 
{D\!\!\!/}_{ch} = i\partial\!\!\!/ - eA\!\!\!/ {(1+i\gamma_5)\over 2}.
\end{equation}
We write $i\Gamma(A) = i\ln{(\det{{D\!\!\!/}_{ch}[A]})}$ so that the effective action 
becomes $S_{eff} = \Gamma[A] +\int -{1\over 4}F^2$. The contribution of the fermionic
loops is contained in the first term of the effective action.

The effective action is not suitable for canonical analysis because of its 
non-local nature
and so is not 
suitable for canonical analysis. It is due to the $\Gamma[A]$, the loop 
contribution and so we introduce an auxiliary bosonic field to express it in terms
of a local action. This can be done by studying the behavior of $\Gamma[A]$ 
under gauge transformation \cite{RAJ, KH1}. The resulting lagrangian is
\begin{eqnarray}
\Gamma[A] &=& 
{1\over 8\pi}\int d^2x tr(N_\mu N^\mu) + 
{1\over 12\pi} \int_v d^3y \epsilon^{ijk}tr(N_i N_j N_k)
 \nonumber  \\  
&+& \int d^2x tr[(-{ie\over 4\pi}(g^{\mu\nu}+\epsilon^{\mu\nu})
N_\mu A_\nu \nonumber \\ 
&+& {e^2\over 8\pi}[(A_0^2-A_1^2) - 2\alpha A_1(A_0+A_1)].
\end{eqnarray}
where $N_\mu=\eta^{-1}
\partial_\mu\eta$. We should mention here that unlike the model presented by Harada in \cite{KH1} this model
lacks manifest Lorentz invariance. 
The first two terms describe the dynamics of the group valued field
$\eta$ and is called the Wess-Zumino action. The first one is the kinetic term and
the second one is the
Wess-Zumino term which we will denote by $S_{wz}$.
The third term
describes its interaction with the gauge
field. The last one is a gauge field mass-like term which is not unique because
of the  ambiguity 
in the regularization. We have taken the generalization of the term that was 
used
in the Abelian model \cite{MG}. In what follows we shall deal with this 
bosonized action.

In order to investigate the consistency of this theory we proceed towards the 
hamiltonian analysis of the theory and determine the phases pace structure. It 
is not difficult  
to calculate the canonical momenta of the various fields. Before that we express
the Wess-Zumino term in the first order form.
\begin{equation}
S_{wz} = {1\over 4\pi} \int d^2x tr[{\cal A}(\eta)\partial_0\eta],
\end{equation}
where ${\cal A}(\eta)$ is not known but the "magnetic field strength" is given
by
\begin{eqnarray}
{\cal F}_{ij;kl} &=& {\partial A_{ij}\over\partial \eta_{lk}} - 
            {\partial A_{kl}\over\partial \eta_{ji}} \nonumber\\
          &=&\partial_1 \eta_{il}^{-1}(x) \eta_{kj}^{-1}(x) - 
             \partial_1 \eta_{kj}^{-1}(x) \eta_{il}^{-1}(x).
\end{eqnarray}

The canonical momenta are obtained by taking the derivative of the action with 
respect to the velocities.
\begin{eqnarray}
p_{ij} &=& {\delta S\over \delta(\partial_0 \eta_{ij})}
         ={1\over 4\pi}[\partial_0\eta_{ji}^{-1} + {\cal A}(\eta)_{ji} - 
          ie ((A_0-A_1)\eta^{-1})_{ji}],\nonumber\\
\Pi^{1a} &=& F^a_{01}. \label{momenta}
\end{eqnarray}
For later convenience  we define 
$\tilde P_{ij} = P_{ji} - {1\over 4\pi}{\cal A}(\eta)_{ij}$.
The hamiltonian is now obtained in straightforward manner
\begin{eqnarray}
H &=& {1\over{8\pi}}\int dx^1 tr(\eta^{1}\partial_0\eta\eta^{-1}\partial_0\eta+    
\eta^{-1}\partial_1\eta \eta^{-1}\partial_1\eta) \nonumber\\ &-& 
    {ie\over 4\pi}\int dx^1 tr \eta^{-1}\partial_1 \eta (A_0-A_1)
   + \int dx^1 tr[{1\over 2}\Pi_1^2 \nonumber\\ 
   &-& {e^2\over 8\pi}(A_0 ^2 - A_1^2+2\alpha A_1(A_0 + A_1)) 
    - (D_1\Pi^1)^aA_0^a] ,
\end{eqnarray}
where we should replace the $\partial_0 \eta$ in terms of the momenta and have 
dropped
a total derivative term.
We can  write 
down the hamiltonian in terms of the free fermionic currents by. 
\begin{eqnarray}
H_c &=& \int dx^1{\cal H}_c\nonumber \\ 
&=& \int dx \pi(\rho_R^2 + \rho_L^2) + e[\rho_R + {e\over 8\pi}(A_0-A_1)](A_0-A_1)\nonumber \\
    & &+ {1\over 2}\Pi^2 - {e^2\over 8\pi}(A_0^2-A_1^2+ 2\alpha A_1(A_0+A_1) 
- (D_1\Pi^1)A_0]. 
\end{eqnarray}
where the fermionic currents are give in terms of canonical variables by
\begin{equation}
\rho_R = {i\over 4\pi}(-4\pi\tilde Pg + g^{-1}\partial_1 g ),~~~~~~ 
\rho_L = {i\over 4\pi}(4\pi g\tilde P - g\partial_1 g^{-1} ),
\label{can} \end{equation}
We find that $\rho_R$ and $\rho_L$ satisfy the following Poisson brackets
 \begin{equation}
\{\rho^a_{R,L}(x), \rho^b_{R,L}(y)\} = -if^{abc}\rho^c_{R,L}(x)\delta(x-y) 
\pm {\delta^{ab}\over 2\pi}\delta'(x-y), \end{equation}
where $\{\rho_R, \rho_L\}$ gives vanishing Poisson brackets.
The Poisson brackets of the group valued  fields are found out to be 
\begin{equation}
\{\eta_{ij}(x), {\tilde P_{kl}}(y)\} = \delta_{ik}\delta_{jl}\delta(x-y),
\end{equation}
\begin{equation}
\{\tilde P_{ij}(x),\tilde P_{kl}(y)\} = -{1 \over 4\pi}{\cal F}(\eta)_{ij,kl}\delta(x-y).
\end{equation}

Let us now introduce the chiral constraint for this non-Abelian version
\begin{equation}
\Omega^a =tr\tau^a\eta {\tilde P^T}+ {1\over {4\pi}}tr\tau^a\partial_1\eta\eta^{-1} \approx 0 \label{CCON}.\end{equation}
If we now impose this constraint in the phases pace of the theory as we have done in the Abelian 
model \cite{AR} we will obtain the hamiltonian in the chiral constraint surface which 
certainly contains  less degrees of freedom. 
\begin{eqnarray}
H_{CH}&=& {1\over {4\pi}}  \int dx^1 tr\eta^{-1}\partial_-\eta\eta^{-1}-{ie\over {2\pi}}\int dx^1tr\eta^{-1}\partial_1\eta(A_0-A_1)
\nonumber\\ 
&-&{e^2\over {4\pi}}\int dx^1 tr((\alpha-1)A_1^2+(\alpha+1) A_1A_0)+
{1\over 2}\int dx^1 (\pi^1)^2 \nonumber \\
&+&\int dx^1 (D_1\pi^1)^a A_0^a. \label{CHAM} \end{eqnarray}
It is straightforward to see that this hamiltonian
can be obtained from the action
\begin{eqnarray}
I_{CH}[\eta, A] &=& {1 \over {4\pi}}\int d^2x tr(\eta^{-1}\partial_1\eta\eta^{-1}\partial_1\eta)
+{1\over {12\pi}}\int d^2x \epsilon^{ijk}N_iN_jN_k \nonumber\\
&-& {ie\over {2\pi}}\int dx^2 tr
\eta^{-1}\partial_1\eta(A_0-A_1) \nonumber\\
&+&{e^2\over{4\pi}}
\int d^2x tr[(\alpha-1)A_1^2 + (\alpha+1)A_1A_0]\nonumber \\
&+&{1\over 2} \int d^2x tr F_{01}^2.\end{eqnarray}

It is the non-Abelian generalization of the action obtained in \cite{AR}. We would also like to
mention that it is the gauged version of the action for the non-Abelian chiral boson with a
generalized Faddeevian regularization. In contrast to the model presented by Harada \cite{KH1} this action too lacks manifest Lorentz covariance.  It is fair to say that the imposition 
of the chiral constraint
in this fashion is not new in field theory. In $(1+1)$ dimensional field theory  Harada is the first person 
to used this \cite{KH}. However 
the scope to use this formalism is very limited. In that sense it is interesting to see the
use of it once more in the non-Abelian case.

Our next task is to quantize the theory. We find that the two primary 
constraints of the theory are
\begin{equation}
\omega_1^a=\pi_0^a = 0,
\end{equation}
\begin{equation}
{\tilde P}_{ij}-{1\over{4\pi}}\partial_1\eta_{ij} =0.
\label{CON}
\end{equation}
The second constraint in (\ref{CON}) reproduces the chiral constraint ({\ref{CCON}). It is convenient 
to write the above constraint multiplying by $\eta$
from the left.
\begin{equation}
(\omega^2)_{ij}=\eta{\tilde P}^T + {1\over{4\pi}}(\partial_1\eta\eta^{-1})_{ij}=0,
\end{equation}
which is equivalent to $\rho_L^a = 0$. It is a second class constraint itself 
having the following Poisson brackets between themselves in two different 
points
 
\begin{equation}
\{\omega^2_{ij}(x), \omega^2_{kl}(y)\}= {1\over{4\pi}}\epsilon_{ij}\delta_{jk}.
\delta(x-y)
\end{equation}

In order to get the full constraint structure we adjoin the primary 
constraints with velocities to the canonical hamiltonian given in (\ref{CHAM}) 
and the total (effective)
hamiltonian turns 
out to be

\begin{equation}
{\cal H}= {\cal H}_c + u^a\rho_L^a + v^a\pi_0^a.
\end{equation}
where ${u^a}$ and ${v^a}$ are velocities corresponding to the primary constraints
The two primary constraints should preserve in time in order to maintain 
consistency of the theory and that leads to two secondary constraints

\begin{equation}
\omega_3^a = (D_1\pi^1)^a + e\rho_R^a +(1+\alpha)
{e^2\over {4\pi}}A_1^a=0,
\end{equation}

\begin{equation}
\omega_4^a=(1+\alpha)\pi^{1a} + 2\alpha(A_0+A_1)'=0,
\end{equation}
and the velocity $u^a$ is found out to be 
\begin{equation}
u^a = -i\eta(\eta^{-1}\partial_1 \eta - ieA_-)\eta^{-1}.
\end{equation}
$\omega_3^a$ is known as the gauss law constraint of the theory. The Poisson brackets between the different constraints of the theory are 

\begin{equation}
\{\omega_1^a(x), \omega_4^b(y)\}=2\alpha\delta'(x-y),
\end{equation}
\begin{equation}
\{\omega_4^a(x), \omega_4^b(y)\}=4\alpha(1+\alpha)\delta'(x-y),
\end{equation}
\begin{equation}
\{\omega_3^a(x), \omega_4^b(y)\}=2\alpha D^{ab}\delta'(x-y) 
+ (1+\alpha) {e^2 \over{4\pi}}\delta(x-y), 
\end{equation}
\begin{equation}
\{\omega_3^a(x), \omega_3^b(y)\} =ef^{abc}[\omega_3^c+{e^2 \over{4\pi}}
(1+\alpha)A_1^c]
 + {e^2 \over{2\pi}}\alpha
\delta^{ab}\delta'(x-y).
\end{equation}
where $D^{ab}\delta(x-y)=
[-\partial_1^a\delta^{ab}+if^{abc}A_1^c(x)]\delta(x-y)$.
Other Poisson brackets give vanishing values. It is now straightforward to 
see that the Dirac brackets  \cite{DIR} of the fields describing the hamiltonian in the
constraint surface is 
\begin{equation}
[A_1^a(x), A_1^b(y)]^*={2\pi\over {e^2}}\delta^{ab}\delta'(x-y),
\end{equation}
\begin{equation}
[A_1^a(x), \pi_1^b(y)]^* = {(\alpha-1)\over {2 \alpha}}\delta^{ab}\delta(x-y),
\end{equation}
\begin{equation}
[\pi_1^a(x), \pi_1^b(y)]^* = {(\alpha+1)^2\over {16\alpha\pi}}
\delta^{ab}\epsilon(x-y).
\end{equation}

We can now impose the constraints of the theory strongly into the 
hamiltonian and find out the hamiltonian on the constrained surface
which will be consistent with the Dirac brackets. 
The hamiltonian density in the reduced space is
\begin{equation}
{\cal H}_R = {\pi \over {e^2}}(D_1\pi^1)^2 +
{1\over 2}(1-\alpha)(D_1\pi^1)^aA^{1a} + {e^2\over {16\pi}}
[(1+\alpha)^2 -8\alpha]A_1^2. \label{RHAM}
\end{equation}

Seeing the bosonized lagrangian one may think that the Lorentz non-invariant 
counter term may spoil the explicit
Lorentz invariance of the lagrangian. However, a closer look reveals the 
invariance
of the theory on the physical subspace. The fact that it is maintained only on
the physical subspace instead of the whole phase space is because of its 
deceptive appearance. In order to show the invariance we have to demonstrate 
the validity of the Poincare' algebra
of this two dimensional system. There are three elements in the algebra, the 
hamiltonian $H$, the momentum $P$ and the boost generator $M$ which have to 
satisfy the relation 
\begin{equation}
\{P, H\} = 0,~~~~~\{M, P\} = -H,~~~~~\{M, H\} = -P,
\label{alz} \end{equation}
to ensure the Poincare' invariance of a theory.
The hamiltonian density has already been evaluated in (\ref{RHAM}).
The momentum density can be written, by discarding a total derivative, as
\begin{eqnarray}
{\cal P}(x) &=& tr\pi^{1a} \partial_1A_1^a + tr\tilde{P}^T\partial_1\eta  \nonumber\\
     &=& \pi^{1a}A_1^a+{\pi\over e^2}[(D_1\pi^1)^a+{(1+\alpha)e^2\over {4\pi}}
A_1^a]^2.
 \label{mom}\end{eqnarray}
The boost generator can be expressed in terms of the hamiltonian and the momentum
densities as 
\begin{equation}
M = tP + \int dx x{\cal H}_R(x).
\end{equation}
From the above expression one can evaluate the left hand sides of (\ref{alz}).
One should 
use the Dirac brackets and that involves a tedious calculation. However, there 
arises a simplification due to the fact that the brackets between the different
terms in the densities are all canonical except the terms appeared as 
$\{A_1(x), A_1(y)\}$ and $\{\pi^1(x), \pi^1(y)\}$.
The rest is a straightforward calculation of canonical brackets which 
yields 
\begin{equation}
\{{\cal H}_R(x), p(y)\} = \partial_1{\cal H}_R(x)\delta(x-y) + ..,~~~ 
\{{\cal H}_R(x), {\cal H}_R(y)\} = \partial_1p(x)\delta(x-y) + ..,
\end{equation}
where the dots represent some $\delta'(x-y)$ terms. Now substituting this 
expressions one can easily check the Poincare' algebra (\ref{alz}). This 
implicit invariance 
of the theory suggests that there may exist a manifestly Lorentz invariant form
of lagrangian \cite{BEL}.

So the non-Abelian generalization of the chiral Schwinger model with new 
Faddeevian regularization \cite{WOT, AR}, as shown above, has a consistent and Lorentz invariant hamiltonian 
structure. The unitarity of the model is not obvious and one has to go through a
BRST analysis for formal proof of unitarity. However we can expect that the theory will respect the unitarity since $QCD_2$ is a super renormalizable theory.
The unraveling of its physical properties would be the next important task.
But unfortunately, unlike its Abelian ancestor, it does not have exact 
solvability. Further, the absence of the gauge invariance does not allow us to 
use the simplifications associated with the light-cone gauge that occurs in
vector $QCD_2$ \cite{BOOK}. The gauge invariance however can be recovered 
by going to its
gauge invariant formulation. But the price to be paid is the extra Wess-Zumino 
field. This formulation can help to study the existence of the mesonic bound states,
and it may also help to study the  different regimes of a theory that are associated with the different extremes of the
coupling and the relation between them. 
The non-Abelian duality of this model \cite{BOOK} which involves writing 
down of a theory in terms of some other fields with the coupling constant inverted is also another interesting thing to be considered seriously. Such studies may help the understanding of non-perturbative physics.

It is a pleasure to thank Prof. P. Mitra for useful discussions. I also want to thank the Director and the 
Head of the Theory Division of Saha Institute of Nuclear Physics, Kolkata for providing computer facilities.

\end{document}